%% file: ARENA2024_A5PA.tex
\documentclass[a4paper,11pt]{article}
\usepackage{pos}
\usepackage{graphicx}
\usepackage{textcomp}
\usepackage{lipsum} 
\usepackage{setspace}

\title{A search for the ultra high energy neutrinos with the low threshold phased array trigger system of the Askaryan Radio Array}
\ShortTitle{Ultra High Energy Neutrino Search with the Phased Array Trigger System in ARA}

\author*[a]{Paramita Dasgupta}
\onbehalf{for the ARA Collaboration \\{\normalsize \normalfont(a complete list of authors can be found at the end of the proceedings)}\\}

\affiliation[a]{Dept. of Physics, Center for Cosmology and AstroParticle Physics,\\
  The Ohio State University, Columbus, OH 43210}

\emailAdd{dasgupta.80@osu.edu}

\abstract{The Askaryan Radio Array (ARA) is an in-ice ultra high energy (UHE, $>10$ PeV) neutrino experiment at the South Pole that aims to detect UHE neutrino-induced radio emission in ice. ARA consists of five independent stations each consisting of a cubical lattice of in-ice antenna clusters with a side length of $\sim$10 m buried at $\sim$200 m below the ice surface. The fifth station of ARA (A5) is special as this station has an additional central string, the phased array (PA), which provides an interferometric trigger that enables ARA to trigger on weak signals that are otherwise buried in noise. Leveraging the low threshold phased array trigger, ARA was the first radio neutrino experiment to demonstrate significant improvement in sensitivity to weak signals. In this contribution, we present initial results from a neutrino search combining information from both the traditional station antennas and the phased array antennas of the A5 station. We show the improved vertex reconstruction achieved with this approach, and leveraging this improvement, we expect to enhance the analysis efficiency beyond what has been achieved previously by ARA. This analysis is the paradigmatic representation of future neutrino searches with the next generation of in-ice neutrino experiments.

}

\FullConference{10th International Workshop on Acoustic and Radio EeV Neutrino Detection Activities (ARENA2024)\\
11-14 June 2024\\
The Kavli Institute for Cosmological Physics, Chicago, IL, USA\\}

\begin{document}
\maketitle
\section{Introduction}\label{sec1:intro}
\par
Over the past three decades, the study of the ultra high energy (UHE, $>$10 PeV) universe has become a significant subfield within particle astrophysics, encompassing both experimental and theoretical research on charged and neutral particles. Ultra high energy cosmic rays (UHECR) with energies greater than 1 EeV are predicted to travel only about 50 Mpc before they interact with Cosmic Microwave Background (CMB) photons and the cosmic ray flux sharply drops at an energy of $\sim$100 EeV, a phenomenon known as the Greisen-Zatsepin-Kuzmin (GZK) cutoff. Due to this abrupt suppression of the cosmic ray flux at Earth~\cite{Greisen:1966jv,Zatsepin:1966jv} to the level of $\sim$1 particle per square kilometer per century, the detection of cosmic rays at the highest energies requires an array of detectors spread over a very large area. Additionally, since cosmic rays are charged particles, their paths are deflected by intergalactic magnetic fields, which scrambles their arrival directions.
\par
In contrast, neutrinos, being chargeless, are unaffected by intergalactic as well as Earth’s magnetic fields. Neutrinos rarely interact with matter and reach us unscathed from the edge of the Universe. Due to these advantages, neutrinos are considered the ideal Standard Model probe of high energy astrophysics at cosmic distances. The GZK suppression of the cosmic ray flux leads to the generation of cosmogenic, or GZK, neutrinos. Ultra high energy neutrino interactions in dense dielectric media produce a burst of broadband radio waves along the Cherenkov cone, known as the ``Askaryan" emission~\cite{Askaryan:1961pfb,ANITA:2006nif}. The attenuation length of radio waves in glacial ice is $\sim$1 km, so the neutrino-induced radio emission travels long distances before being absorbed in the ice. This allows instrumentation to be placed relatively sparsely over a large area of roughly $100$~km$^2$, required to detect the small neutrino flux with energy beyond $\sim$10 PeV. 
\par 
The Askaryan Radio Array (ARA) at the South Pole is the longest-running in-ice radio experiment that is specifically designed to detect the UHE neutrino-induced Askaryan emission in ice. In this proceeding, we report on the progress towards a search for UHE neutrinos using the newest ARA station, which features a beamforming trigger system designed to capture weak radio signals.
\section{The ARA Detector }\label{sec2:detector}
\par
ARA consists of five independent stations, deployed on a hexagonal grid with a 2~km spacing between neighboring stations, to maximize the effective area at $E_{\nu}>10^{18}$~eV. Each ARA station includes four ``measurement strings'' deployed in vertical boreholes drilled to a depth of $\sim$200~meter. Each string has two vertically-polarized (VPol) and two horizontally-polarized (HPol) antennas. Both the VPol and HPol antennas are sensitive to radio frequency (RF) signals from $150-850$~MHz~\cite{ARA:2019wcf}. 
\par
In addition to the receiving antennas, each ARA station is equipped with two calibration strings, deployed at the same depth as the four measurement strings, and typically located $\sim$40 meters from the center of the station. Each calibration string contains one HPol and one VPol transmitter antenna, both capable of emitting broadband RF pulses or RF noise for in-situ geometry calibration. ARA stations maintain an approximate $6$ Hz trigger rate, which includes a $1$ Hz calibration pulser rate for in-situ antenna geometry and timing calibration, as well as an additional $1$ Hz forced trigger rate to monitor ambient noise and detector performance.

\begin{figure}[htbp!]
  \centering
  \includegraphics[width=0.45\textwidth]{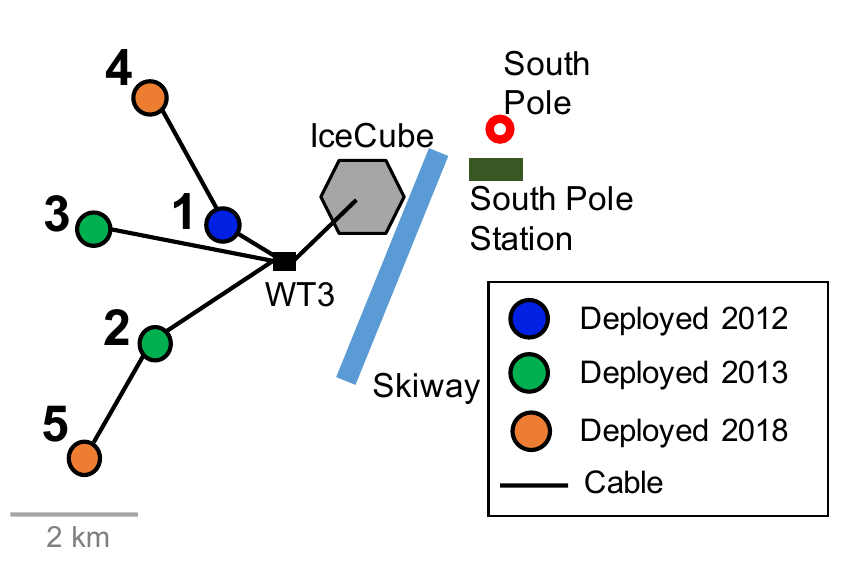}
  \hfill
  \includegraphics[width=0.45\textwidth]{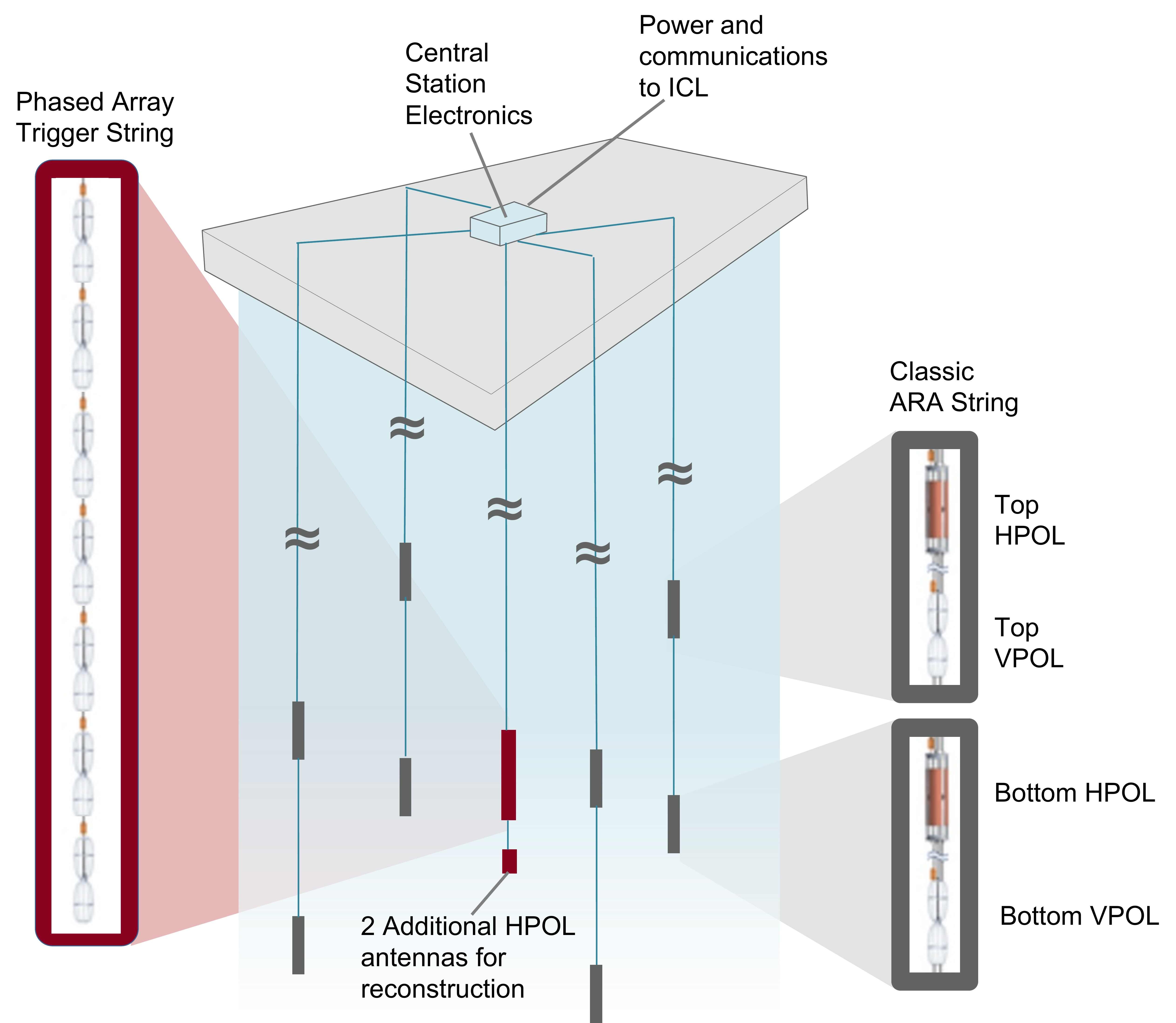}
  \caption{Left: Layout of the ARA station array along with the deployment timeline. Right: Layout of the two subdetectors of ARA's fifth station, A5. The traditional ARA antennas are shown (grey). Except for the central phased array string with closely packed antennas (red), all other ARA stations (A1-A4) have the same layout as the A5 station.}
  \label{fig:A5_detector}
\end{figure}
\subsection{The Phased Array Trigger Design}\label{sec3:pa_trigger}
\par
The fifth station of ARA (A5) is special since this is the only station of ARA that is equipped with two subdetectors: the traditional ARA detector, described in the previous section, and an additional ``phased array'' (PA) detector (see Fig.~\ref{fig:A5_detector}, right panel). The PA consists of a single string with nine closely spaced antennas (seven VPols, and two HPol antennas), deployed at a depth of ${\sim}180$~m. This hybrid configuration comprising traditional measurement antennas and the phased array antennas (see Fig.~\ref{fig:A5_detector}, right panel) has informed the design of the next-generation UHE neutrino detectors, including RNO-G~\cite{RNO-G:2020rmc} and the radio component of IceCube-Gen2~\cite{IceCube-Gen2:2020qha}. 
\par
Each of these subdetectors has its data acquisition (DAQ) system and trigger system~\cite{ARA:2021wmr}. The PA uses an interferometric trigger that sums waveforms in the 7~VPol channels using inter-channel timing offsets based on a plane wave signal from one of $15$ predefined zenith directions, called ``beams''.
Impulsive radio signals coming from one of the predefined directions add approximately coherently, while noise adds incoherently in all beams. The PA instrument triggers when the power in a beam exceeds the threshold power in a $10$~ns window and maintains an $11$~Hz global trigger rate across all beams. This trigger strategy significantly improves the trigger efficiency, in particular for low signal-to-noise ratio (SNR) signals~\cite{Allison:2018ynt}. 
\par 
Initially, the traditional ARA subdetector and the PA subdetector (see Fig.~\ref{fig:A5_detector}, right panel) were two independent detectors. However, in 2020, the traditional ARA DAQ was lost due to a USB port failure. To solve this issue, six ARA VPol channels were directly connected to the PA DAQ's six available channels. This combined system is referred to as the A5/PA hybrid system to distinguish it from when both systems operated independently. This configuration also serves as a model for the hybrid DAQ systems planned for next-generation UHE neutrino detectors. The analysis presented in this proceeding is based on the data collected by this merged system over two years ($2020$-$2021$). 
\section{Simulation}\label{sec3:simu}
\par
We use ARA's standard Monte Carlo event generation and detector simulation package \verb|AraSim|~\cite{ARA:2014fyf} to simulate a common set of events in both the PA and the traditional ARA DAQs. The ARA collaboration has significantly improved the accuracy and reliability of its simulation by incorporating several changes. One of the key improvements in simulation was done by generating a data-driven noise model for all ARA stations. For this A5/PA hybrid analysis, we developed a data-driven noise model by analyzing the forced trigger events in each antenna.
Secondly, new anechoic chamber measurements of frequency-dependent gain pattern for three categorizations of ARA antennas (HPols, quadslot antennas; top and bottom VPols, dipole antennas) were taken. Furthermore, we developed a data-driven signal chain amplitude gain model by incorporating a realistic noise model for each antenna, thermal noise from ambient ice temperature, and the properties of the antenna \& amplifiers. 
\par
The simulated events are generated according to a $\phi \propto E^{-1}$ power-law spectrum and are assigned a uniformly distributed source direction and vertex locations within 8~km of the detector. 
The PA trigger is calculated based on an estimation that is accurate within 15\%: the observed SNR of an event and the difference between the signal's emission angle compared to the Cherenkov angle are used to estimate the trigger efficiency of each event. If a random number between $0$ and $1$ is equal to or less than the calculated trigger efficiency at a given signal-to-noise ratio (SNR), the event is said to have been triggered. 
\section{Pioneering Hybrid Analysis}\label{sec3}
\par
A previous search for diffuse neutrinos with the PA instrument alone demonstrated both the improved trigger efficiency and analysis efficiency possible using an interferometric trigger~\cite{ARA:2022rwq}.
\begin{figure}[htbp!]
  \centering
    \includegraphics[width=0.90\textwidth]{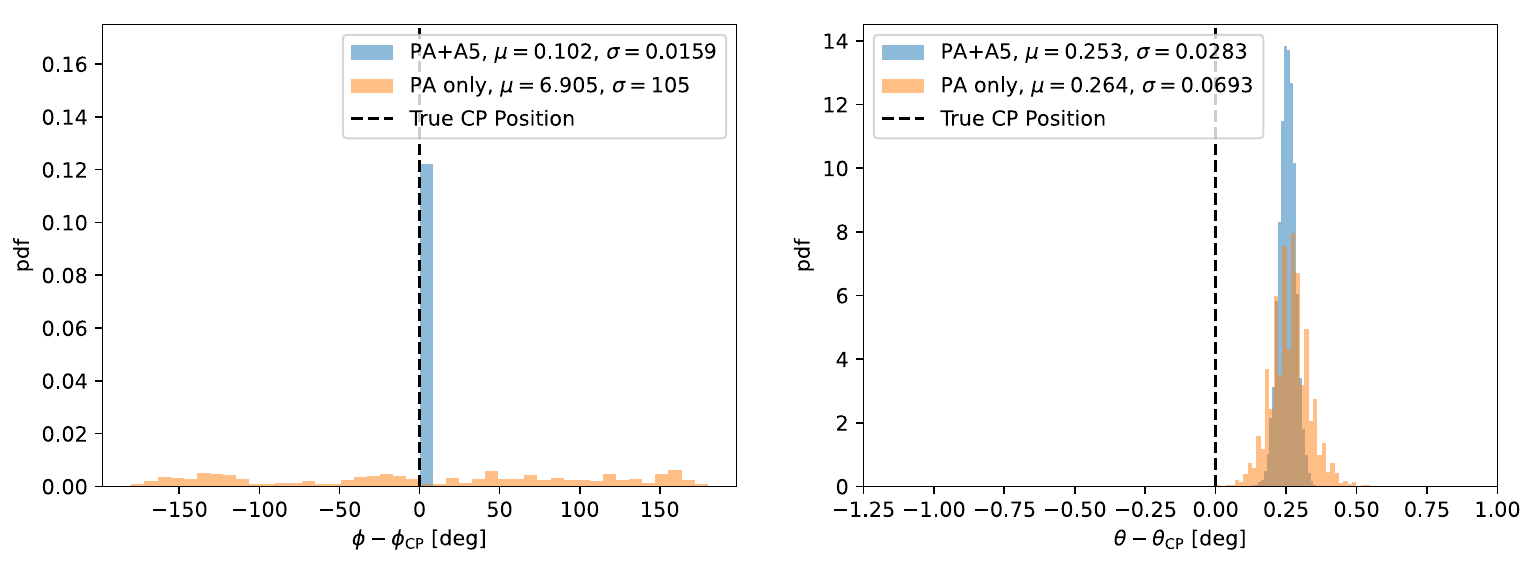}
    \caption{Reconstructed $\phi$ (azimuth angle, left panel) and $\theta$ (zenith angle, right panel) of the calibration pulser location in ice relative to the true pulser location ($\phi_{CP}$, $\theta_{CP}$, dashed line). Distributions are shown for the A5/PA hybrid system (blue) and the PA system alone (orange).}
    \label{fig:a5paReco}
    \vspace{-5pt}
\end{figure}
In this analysis, we use the hybrid design to demonstrate the additional benefit of using traditional antennas from the ARA subdetector (see Fig.~\ref{fig:A5_detector}, right panel). We expect this hybrid configuration to have two main benefits. The first benefit is the excellent azimuth reconstruction of events using the antennas from both the ARA and PA subdetector (see Fig.~\ref{fig:A5_detector}, right panel). Since the PA instrument consists of a single string of 9 closely spaced antennas, we do not have azimuth information for the events in our dataset (see Fig.~\ref{fig:a5paReco}, left panel). Due to the azimuthal symmetry of the PA instrument, it has been a challenge to remove the background in a previous analysis with the PA instrument alone~\cite{ARA:2022rwq}. In contrast to PA antennas, ARA traditional antennas provide us with precise and accurate azimuth angle information and $\sim$2 times better accuracy in zenith angle information for high SNR events in our dataset (see Fig.~\ref{fig:a5paReco}). This major improvement in event reconstruction allows for more efficient identification of sources of impulsive anthropogenic events.
The second benefit of having a hybrid design lies in the measurement of radio signals in more channels. This allows for improved discrimination power between physics events and background events than is possible with the phased array antennas alone.
\section{Data Analysis}
\par
This analysis uses $504$ days of data taken over the years 2020 and 2021, a period that spans after ARA DAQ was lost and the ARA channels were connected to the PA subdetector's open channels. 
This analysis is performed in a blind fashion~\cite{Klein:2005di} where we generate a sample dataset, which is 10\% of the total data. This sample dataset is generated by randomly selecting one event out of every ten events so that the entire livetime of this analysis is sampled and any time-dependent noisy period is well represented in the sample dataset. We analyze this sample dataset to develop quality cuts to discriminate a possible neutrino signal against the known sources of background. The remaining 90\% of the total data will be used to set the diffuse neutrino flux limit.

\subsection{Correlation Mapping}\label{sec5:corr_map}
\par
\begin{figure}[htbp!]
  \centering
  
  {\includegraphics[width=0.540\textwidth]{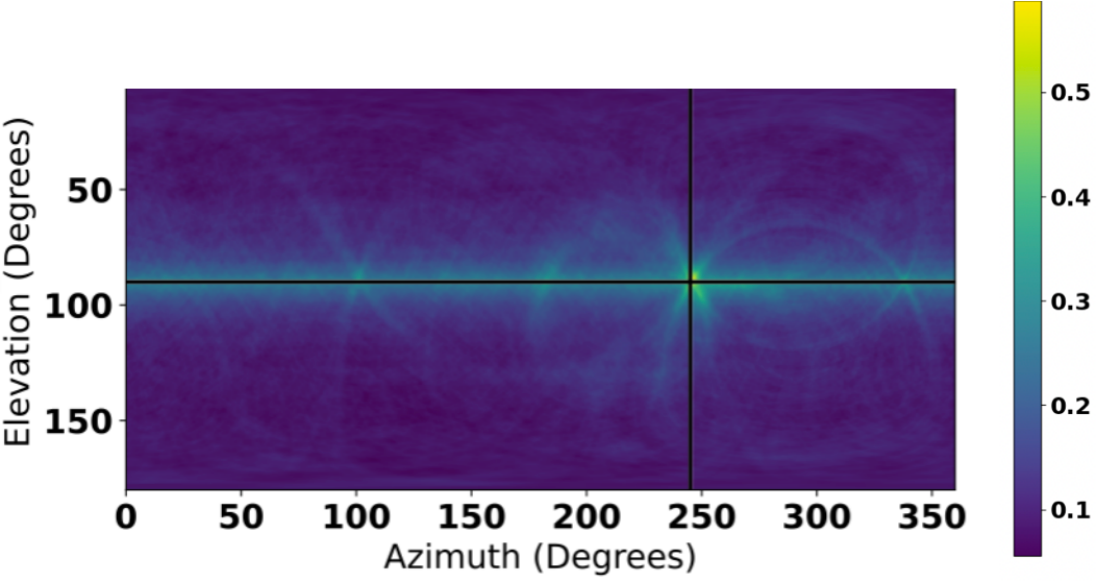}}
  \hfill
  {\includegraphics[width=0.40\textwidth]{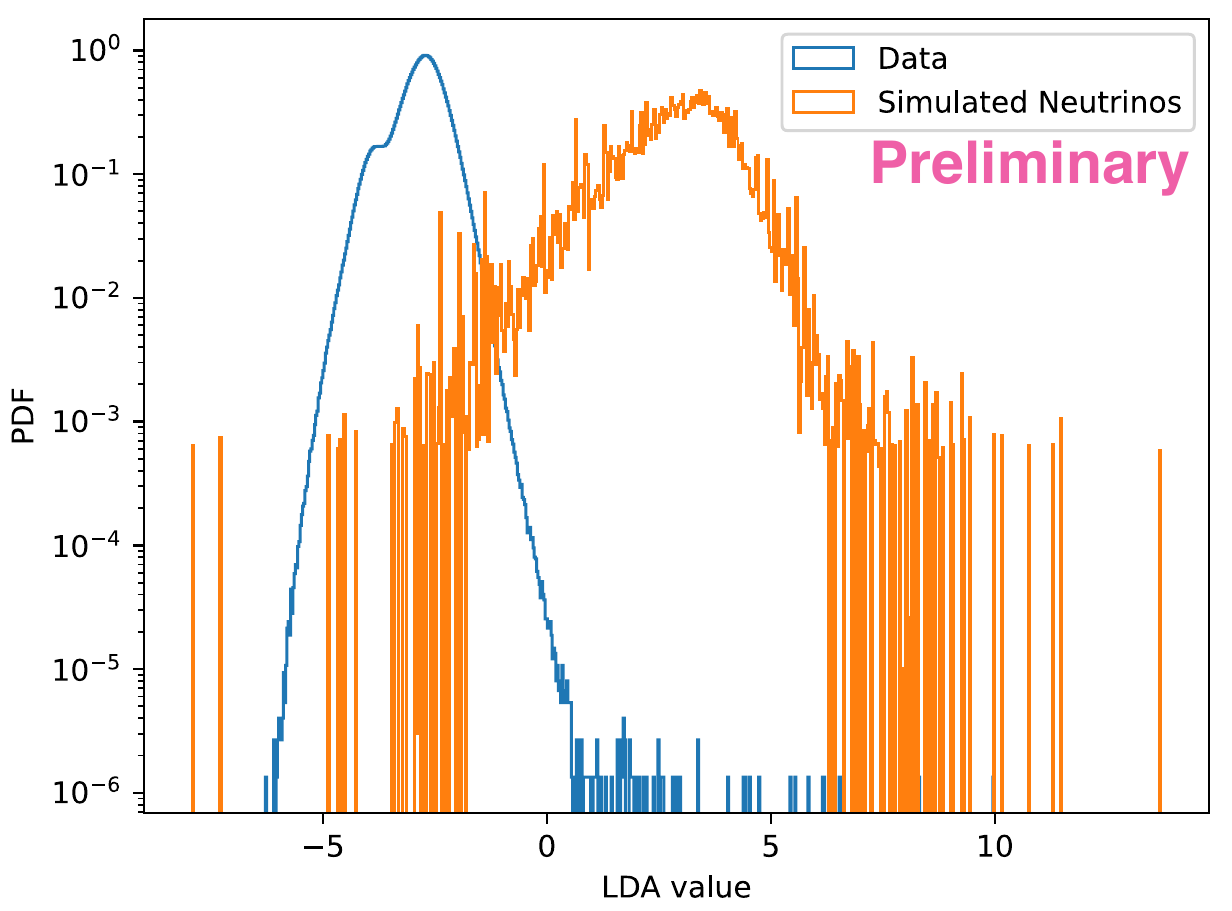}}
  \caption{Left: Correlation map for a calibration event, recorded by A5/PA hybrid system. The horizontal and vertical lines (black) correspond to the true elevation angle (where, elevation angle $=90^\circ-$ zenith angle) and azimuth angle of the calibration event's source location respectively from the center of the A5/PA hybrid detector. Right: Preliminary distribution of linear discriminant value for simulated neutrinos (orange) and thermal noise dominated events (blue) in the 10\% sample dataset.}
  \label{fig:lda}
  \vspace{-5pt}
\end{figure}

One of the important tools for this analysis is two-dimensional reconstruction in the zenith ($\theta$) and azimuth ($\phi$) direction from the phased array string. For each event in our 10\% sample dataset, we generate a 2D correlation map (see Fig.~\ref{fig:lda}, left panel) for a given radius ($R$) from the phased array string. 

For each point on the ($\theta$-$\phi$) correlation map, we add the signals on each channel of the A5/PA combined system, as if the signals are coming from that point on the map and calculate the average value of the sum across all channel pairs for each ($\theta$-$\phi$) point on the map. The brightest point on the map is the most likely origin of the signal (see Fig.~\ref{fig:lda}, left panel). We find the time delay corresponding to the brightest peak on the correlation map and add the signals together using that time delay to form a coherently summed waveform (CSW) for each event in our data. From the correlation map and the CSW, we derive a list of analysis variables such as maximum correlation value, SNR of the CSW, SNR of the Hilbert envelope of the CSW, impulsivity, reconstructed $\theta$ \& $\phi$ (see Fig.~\ref{fig:lda}, left panel) for each event in our dataset. 
Each of these variables describes various characteristics of the events in our data and has excellent signal-background discrimination power.

\subsection{Background}\label{sec:backgrounds}
The main sources of background in this analysis are thermal noise, anthropogenic, and cosmic ray events. We remove the known calibration data (see Section~\ref{sec3:pa_trigger}) using the timestamp of the data and by implementing an additional geometric cut in a one square degree area around the calibration pulser location (to remove any mistimed pulses). In addition to calibration data, we also have forced trigger events due to the self-triggering of both subdetector DAQs every second. These events are expected to be random samples of noise, and hence they are also removed from our dataset using a forced trigger tag. Importantly, by removing these tagged events, we are not able to remove all thermal noise events from our dataset, as upward fluctuations of the noise can satisfy the PA trigger condition occasionally.

\par
The next largest background we have in our data is the continuous wave (CW) events at known frequencies of ${\sim}410$~MHz, from the weather balloon launched twice every day at the South Pole, and ${\sim}210$~MHz, from satellite communications. We apply a sine-subtraction CW filter, first implemented by ANITA Collaboration~\cite{ANITA:2018vwl} to remove the CW contamination from our data.  
\par
The final backgrounds in our data are the most challenging to remove: impulsive anthropogenic events and cosmic ray events. Both of these event types have waveforms that are similar to those expected for neutrino events, so careful removal of these backgrounds has a significant impact on the efficiency of this analysis. 
Impulsive anthropogenic events are expected to originate from above the ice surface and, possibly, reconstruct to a known source (e.g. South Pole Station). Additionally, one expects that such events should be clustered temporally and spatially. Hence, a spatio-temporal cut will help us remove them from our dataset. We developed a spatio-temporal-clustering algorithm to identify and remove these backgrounds. Cosmic ray events are also expected to originate from above the ice surface. In particular, cosmic rays induce particle showers, both in the atmosphere and in the ice, which emit radio pulses very similar to those predicted for neutrino-initiated particle showers~\cite{PhysRevD.110.023010}. Using an expected zenith distribution of the cosmic ray events, we apply a cut on the zenith distribution of events to remove the cosmic ray events from our dataset. 

\subsection{Fisher Discriminant modeling }\label{sec5:Fisher}
After we remove the non-thermal background from the 10\% sample, the remaining events are expected to be dominated by thermal noise fluctuation. To maximize the separation between the simulated neutrino signals and the thermal noise-dominated events, we perform a linear discriminant analysis (LDA) against the remaining sample and the simulated neutrino events (see Section~\ref{sec3:simu}). 
The LDA analysis uses a set of selection variables such as CSW SNR, maximum correlation, the impulsivity of the signal, etc (see Section~\ref{sec5:corr_map}) and combines these variables into a single, final selection variable called LDA value (see Fig.~\ref{fig:lda}, right panel). A full description of these analysis variables is beyond the scope of this proceeding. Our next step is to optimize the location of the cut on LDA value for best-expected sensitivity, with the sensitivity defined as the best 90\% Feldman-Cousins upper limit~\cite{PhysRevD.57.3873}  

\section{Conclusion}
\par
Over the past six years, the ARA Collaboration has successfully demonstrated the advantages of a low-threshold trigger analysis for the radio detection of UHE neutrinos using its phased array detector. In this study, we outline a pioneering analysis that combines data from a hybrid detector system comprising traditional outer antennas and a low-threshold phased array beamforming instrument, and we present the first results. 
\begin{figure}[htbp!]
  \centering
  \includegraphics[width=0.5500\textwidth]{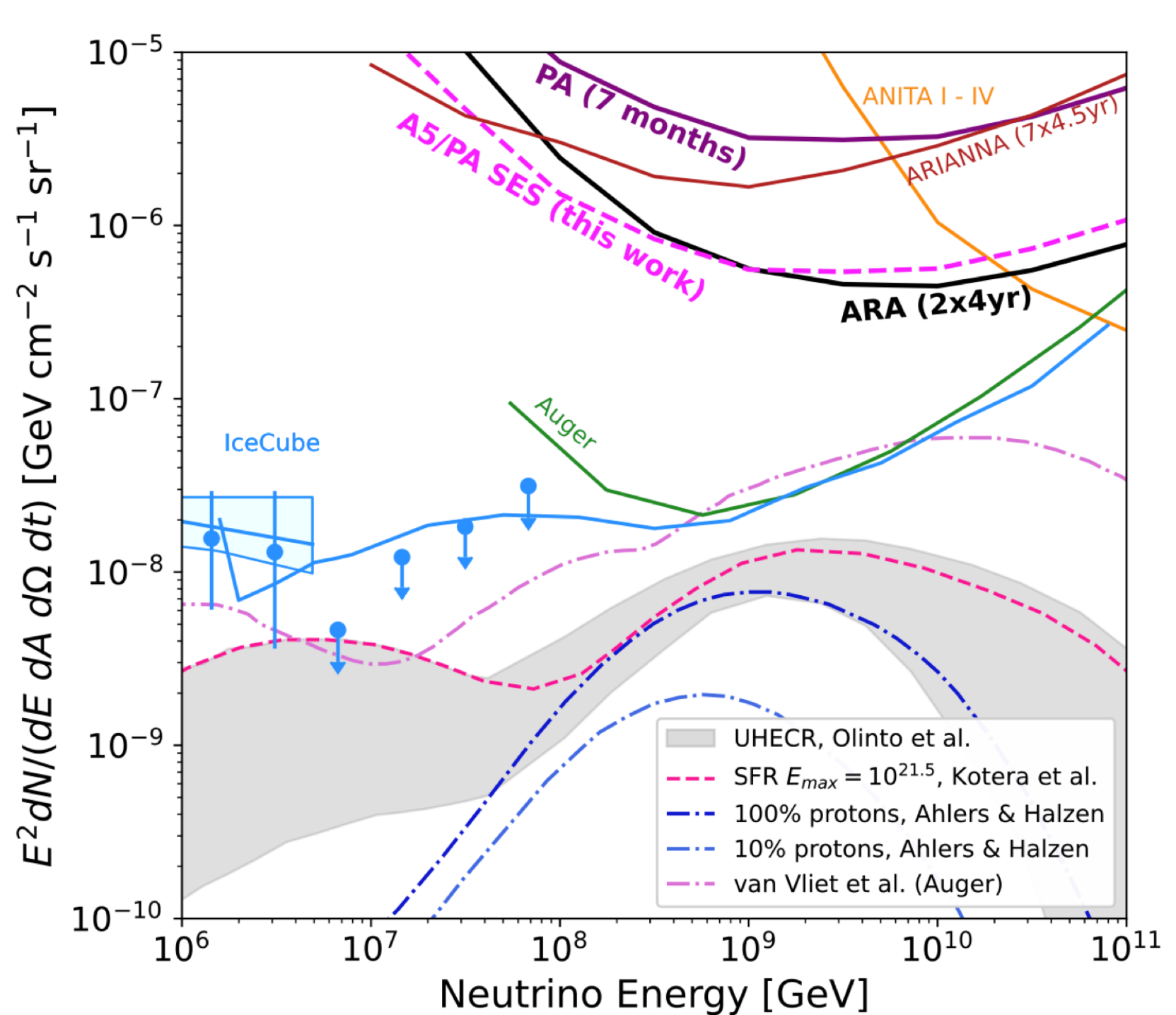}
  \caption{Projected single-event sensitivity of the diffuse neutrino flux at analysis level from the A5/PA hybrid analysis (dashed), compared to previous ARA analyses. Data, limits, theoretical predictions, and sensitivities of other experiments are also shown for reference.} 
  \label{fig:sensitivity}
\end{figure}

\par
This analysis is first-of-its-kind and a crucial advancement in improving ARA's sensitivity to energies below 1 EeV (see Fig.~\ref{fig:sensitivity}). The expected neutrino flux limit of the phased array instrument alone improves quickly with the increased livetime due to the highly efficient trigger of the phased array. This improvement at the trigger level is translated to the analysis efficiency of the phased array. Using a hybrid detector system, which includes both the phased array and the traditional antennas, we expect that the trigger efficiency and analysis efficiency can be further improved.
We are optimistic about the use of this hybrid design in the future. Also, the developed analysis described in this paper will be instrumental in developing new tools for next-generation radio observatories, such as RNO-G and the radio component of IceCube-Gen2. Simultaneously, the ARA Collaboration has embarked on the ambitious task of analyzing the entirety of the collected data from the full array~\cite{ARA:2023frx}. With an accumulated dataset of approximately 24 station-years, ARA has the opportunity for groundbreaking discoveries and the power to constrain cosmogenic models.


\bibliographystyle{JHEP}
\bibliography{references}

%

\clearpage

\input{ara_arena_authors}

\input{ara_arena_acknowledgements}

\end{document}

%% file: ara_arena_authors.tex
\section*{Full Author List: ARA Collaboration (July 22, 2024)}

\noindent
S.~Ali\textsuperscript{1},
P.~Allison\textsuperscript{2},
S.~Archambault\textsuperscript{3},
J.J.~Beatty\textsuperscript{2},
D.Z.~Besson\textsuperscript{1},
A.~Bishop\textsuperscript{4},
P.~Chen\textsuperscript{5},
Y.C.~Chen\textsuperscript{5},
Y.-C.~Chen\textsuperscript{5},
B.A.~Clark\textsuperscript{6},
A.~Connolly\textsuperscript{2},
K.~Couberly\textsuperscript{1},
L.~Cremonesi\textsuperscript{7},
A.~Cummings\textsuperscript{8,9,10},
P.~Dasgupta\textsuperscript{2},
R.~Debolt\textsuperscript{2},
S.~de~Kockere\textsuperscript{11},
K.D.~de~Vries\textsuperscript{11},
C.~Deaconu\textsuperscript{12},
M.~A.~DuVernois\textsuperscript{4},
J.~Flaherty\textsuperscript{2},
E.~Friedman\textsuperscript{6},
R.~Gaior\textsuperscript{3},
P.~Giri\textsuperscript{13},
J.~Hanson\textsuperscript{14},
N.~Harty\textsuperscript{15},
K.D.~Hoffman\textsuperscript{6},
M.-H.~Huang\textsuperscript{5,16},
K.~Hughes\textsuperscript{2},
A.~Ishihara\textsuperscript{3},
A.~Karle\textsuperscript{4},
J.L.~Kelley\textsuperscript{4},
K.-C.~Kim\textsuperscript{6},
M.-C.~Kim\textsuperscript{3},
I.~Kravchenko\textsuperscript{13},
R.~Krebs\textsuperscript{8,9},
C.Y.~Kuo\textsuperscript{5},
K.~Kurusu\textsuperscript{3},
U.A.~Latif\textsuperscript{11},
C.H.~Liu\textsuperscript{13},
T.C.~Liu\textsuperscript{5,17},
W.~Luszczak\textsuperscript{2},
K.~Mase\textsuperscript{3},
M.S.~Muzio\textsuperscript{8,9,10},
J.~Nam\textsuperscript{5},
R.J.~Nichol\textsuperscript{7},
A.~Novikov\textsuperscript{15},
A.~Nozdrina\textsuperscript{1},
E.~Oberla\textsuperscript{12},
Y.~Pan\textsuperscript{15},
C.~Pfendner\textsuperscript{18},
N.~Punsuebsay\textsuperscript{15},
J.~Roth\textsuperscript{15},
A.~Salcedo-Gomez\textsuperscript{2},
D.~Seckel\textsuperscript{15},
M.F.H.~Seikh\textsuperscript{1},
Y.-S.~Shiao\textsuperscript{5,19},
S.C.~Su\textsuperscript{5},
S.~Toscano\textsuperscript{20},
J.~Torres\textsuperscript{2},
J.~Touart\textsuperscript{6},
N.~van~Eijndhoven\textsuperscript{11},
G.S.~Varner\textsuperscript{21,$\dagger$},
A.~Vieregg\textsuperscript{12},
M.-Z.~Wang\textsuperscript{5},
S.-H.~Wang\textsuperscript{5},
S.A.~Wissel\textsuperscript{8,9,10},
C.~Xie\textsuperscript{7},
S.~Yoshida\textsuperscript{3},
R.~Young\textsuperscript{1}
\\
\\
\textsuperscript{1} Dept. of Physics and Astronomy, University of Kansas, Lawrence, KS 66045\\
\textsuperscript{2} Dept. of Physics, Center for Cosmology and AstroParticle Physics, The Ohio State University, Columbus, OH 43210\\
\textsuperscript{3} Dept. of Physics, Chiba University, Chiba, Japan\\
\textsuperscript{4} Dept. of Physics, University of Wisconsin-Madison, Madison,  WI 53706\\
\textsuperscript{5} Dept. of Physics, Grad. Inst. of Astrophys., Leung Center for Cosmology and Particle Astrophysics, National Taiwan University, Taipei, Taiwan\\
\textsuperscript{6} Dept. of Physics, University of Maryland, College Park, MD 20742\\
\textsuperscript{7} Dept. of Physics and Astronomy, University College London, London, United Kingdom\\
\textsuperscript{8} Center for Multi-Messenger Astrophysics, Institute for Gravitation and the Cosmos, Pennsylvania State University, University Park, PA 16802\\
\textsuperscript{9} Dept. of Physics, Pennsylvania State University, University Park, PA 16802\\
\textsuperscript{10} Dept. of Astronomy and Astrophysics, Pennsylvania State University, University Park, PA 16802\\
\textsuperscript{11} Vrije Universiteit Brussel, Brussels, Belgium\\
\textsuperscript{12} Dept. of Physics, Enrico Fermi Institue, Kavli Institute for Cosmological Physics, University of Chicago, Chicago, IL 60637\\
\textsuperscript{13} Dept. of Physics and Astronomy, University of Nebraska, Lincoln, Nebraska 68588\\
\textsuperscript{14} Dept. Physics and Astronomy, Whittier College, Whittier, CA 90602\\
\textsuperscript{15} Dept. of Physics, University of Delaware, Newark, DE 19716\\
\textsuperscript{16} Dept. of Energy Engineering, National United University, Miaoli, Taiwan\\
\textsuperscript{17} Dept. of Applied Physics, National Pingtung University, Pingtung City, Pingtung County 900393, Taiwan\\
\textsuperscript{18} Dept. of Physics and Astronomy, Denison University, Granville, Ohio 43023\\
\textsuperscript{19} National Nano Device Laboratories, Hsinchu 300, Taiwan\\
\textsuperscript{20} Universite Libre de Bruxelles, Science Faculty CP230, B-1050 Brussels, Belgium\\
\textsuperscript{21} Dept. of Physics and Astronomy, University of Hawaii, Manoa, HI 96822\\
\textsuperscript{$\dagger$} Deceased\\

%% file: ara_arena_acknowledgements.tex
\section*{Acknowledgements}

\noindent
The ARA Collaboration is grateful to support from the National Science Foundation through Award 2013134.
The ARA Collaboration
designed, constructed, and now operates the ARA detectors. We would like to thank IceCube, and specifically the winterovers for the support in operating the
detector. Data processing and calibration, Monte Carlo
simulations of the detector and of theoretical models
and data analyses were performed by a large number
of collaboration members, who also discussed and approved the scientific results presented here. We are
thankful to Antarctic Support Contractor staff, a Leidos unit 
for field support and enabling our work on the harshest continent. We thank the National Science Foundation (NSF) Office of Polar Programs and
Physics Division for funding support. We further thank
the Taiwan National Science Councils Vanguard Program NSC 92-2628-M-002-09 and the Belgian F.R.S.-
FNRS Grant 4.4508.01 and FWO. 
K. Hughes thanks the NSF for
support through the Graduate Research Fellowship Program Award DGE-1746045. A. Connolly thanks the NSF for
Award 1806923 and 2209588, and also acknowledges the Ohio Supercomputer Center. S. A. Wissel thanks the NSF for support through CAREER Award 2033500.
A. Vieregg thanks the Sloan Foundation and the Research Corporation for Science Advancement, the Research Computing Center and the Kavli Institute for Cosmological Physics at the University of Chicago for the resources they provided. R. Nichol thanks the Leverhulme
Trust for their support. K.D. de Vries is supported by
European Research Council under the European Unions
Horizon research and innovation program (grant agreement 763 No 805486). D. Besson, I. Kravchenko, and D. Seckel thank the NSF for support through the IceCube EPSCoR Initiative (Award ID 2019597). M.S. Muzio thanks the NSF for support through the MPS-Ascend Postdoctoral Fellowship under Award 2138121. A. Bishop thanks the Belgian American Education Foundation for their Graduate Fellowship support.